\documentclass[aps,prb,twocolumn,superscriptaddress,footinbib]{revtex4-1}
\usepackage{graphicx}
\usepackage{feynmp}
\usepackage{epsfig}
\usepackage[latin1]{inputenc}
\usepackage{amsmath}
\usepackage{amsfonts}
\usepackage{amssymb}
\usepackage{amsmath}
\usepackage{amssymb}
\usepackage{color}
\usepackage{multirow,bigdelim}
\usepackage{enumerate}

\font\elevenmib=cmmib10 scaled 1095
\font\tenmib=cmmib10
\font\eightmib=cmmib10 scaled 800
\font\sixmib=cmmib10 scaled 667
\skewchar\elevenmib='177
\newfam\mibfam

\textfont\mibfam=\tenmib
\scriptfont\mibfam=\eightmib
\scriptscriptfont\mibfam=\sixmib

\begin{document}

\title{A Bionic Coulomb Phase on the Pyrochlore Lattice}

\author{V. Khemani}
\affiliation{Department of Physics, Princeton University, Princeton, New Jersey 08544, USA}
\author{R. Moessner}
\affiliation{Max-Planck-Institut f\"{u}r Physik komplexer Systeme, D-01187 Dresden, Germany}
\author{S. A. Parameswaran}
\affiliation{Department of Physics, University of California, Berkeley, CA 94720, USA}
\author{S. L. Sondhi}
\affiliation{Department of Physics, Princeton University, Princeton, New Jersey 08544, USA}

\date{\today}

\begin{abstract}
A class of three dimensional classical lattice systems with macroscopic ground state degeneracies, most
famously the spin ice system, are known to exhibit ``Coulomb'' phases wherein long wavelength correlations
within the ground state manifold are described by an emergent Maxwell electrodynamics. We discuss a
new example of this phenomenon---the four state Potts model on the pyrochlore lattice---where the long
wavelength description now involves three independent gauge fields as we confirm via simulation.
The excitations above the ground state manifold are bions, defects that are simultaneously charged under
two of the three gauge fields, and exhibit an entropic interaction dictated by these charges. We also show
that the distribution of flux loops 
shows a scaling with loop length and system size previously
identified as characteristic of Coulomb phases.
\end{abstract}
\maketitle
\section{Introduction}
Much recent activity in condensed matter physics has involved the study of ``topologically ordered'' phases which characteristically exhibit emergent gauge fields and deconfined fractionalized excitations at low energies. Canonical quantum examples of these are the various fractional quantum Hall phases \cite{Wen,ZhangKivelson} and much of the physics is present in the elegant classical physics of the spin ice compounds \cite{SLSspinice}.

Gauge fields are intimately connected to local constraints, as in the textbook example of Maxwell
electromagnetism wherein Gauss's law reflects a constraint at each point in space that must be
obeyed by the dynamics. In a condensed matter setting, the analogous constraints arise as a low-energy effective truncation of the space of configurations; examples range
from the dimer configurations
of short range RVB theory \cite{Moessnerdimer} to the string nets of Levin and Wen \cite{LW}.

A subclass of these constraints literally take the form of lattice versions of the familiar Gauss's law
for abelian gauge fields, albeit with restricted microscopically realizable values for the lattice
electromagnetic fields. The introduction of an appropriate classical statistical mechanics that consists
of averaging over all allowed configurations with uniform weight leads to the so-called Coulomb phase\cite{Henleyreview}
with dipolar correlations, whose coarse-grained description realizes a Euclidean Maxwell theory. More
elaborately, the introduction of a quantum dynamics in the constrained manifold can lead to a version of
Maxwell electrodynamics coupled to electric charges and magnetic monopoles. An elegant point of intersection
between the classical and quantum Coulomb phases is an appropriate Rokhsar-Kivelson point where the
ground state wavefunction is itself an equal amplitude superposition of allowed configurations\cite{Hermele04, MoessnerSondhi}.

In this paper we expand the catalog of Coulomb phases. We study the antiferromagnetic four-state Potts model
on the highly frustrated pyrochlore lattice and show that its ground state manifold exhibits correlations
characterized by {\it three} abelian gauge fields. We find that the fundamental excitations/defects above
this ground state manifold are charged under these gauge fields in an unusual way---they carry nonzero charges
for two of the three gauge fields whence we refer to them as bions. (The term dyon is already reserved for
particles charged under dual electric and magnetic fields whereas 
here both fields are of the same species.) In the classical setting, which is our primary interest in this paper, the import of the charge assignments is
that it predicts the entropic force between different bions and more generally the free energy/entropy
for any configuration of multiple bions. Our evidence for these assertions comes from a Monte Carlo simulation
that agrees with the correlations predicted by the triple Maxwell theory, and which yields statistics of flux-loops
in the ground state manifold that have been previously suggested to be a sharp diagnostic of the Coulomb
phase \cite{Jaubert2011}.

Readers familiar with the existing literature on Coulomb phases will note that it is already known \cite{Isakov2004, Henley2005} that
classical $O(N)$ spins with a nearest neighbor antiferromagnetic interaction on the pyrochlore lattice
\begin{enumerate}[(a)]
  \item exhibit a Coulomb phase with one gauge field for $N=1$ (Ising) spins which is the case relevant to spin ice and indeed observed in experiments,
  \item exhibit order by disorder for $N=2$ and
  \item exhibit a Coulomb phase with $N$ gauge fields for $N~\ge~3$.
\end{enumerate}
It is interesting to situate the current work in this context. To this end imagine starting with $O(3)$
symmetric Heisenberg spins that live on the sphere (Fig. \ref{fig:spinspace}c) whose ground state correlations are governed
by three independently fluctuating gauge fields. Excitations above this manifold are gapless and involve
arbitrarily small charges under the gauge fields. If we restrict their range by generating an easy axis
(Fig. \ref{fig:spinspace}a)
we return to the Ising case where the number of gauge fields is now down to one and the excitations are
gapped and quantized. The import of this current paper is that if we restrict the range to four symmetric
points on the sphere (Fig. \ref{fig:spinspace}b) the number of gauge fields is unchanged although the excitations again
become gapped and quantized. We believe that this reduction can be extended to higher dimensions by
considering generalizations of kagome/pyrochlore\cite{ZacharyTorquato} involving 
$d+1$-simplices in $d$ dimensions and starting
with $O(d)$ spins and restricting them to $d+1$ state Potts configurations.

We would be remiss if we did not note that this paper generalizes the early results of Kondev and Henley \cite{Kondev} from the two dimensional lattice known variously as the square lattice with crossings or planar
pyrochlore, to three dimensions. Readers who peruse the early paper will spot the family resemblance immediately.

In the balance of the paper, we will set up the Potts model and its mapping to vector spins
(Section II), map these in turn to a coarse grained description in terms of pseudo-magnetic flux/gauge fields and
confirm the resulting predictions for the correlations (Section III), discuss the bionic
excitations (Section IV), study the statistics of loops (Section V) and conclude with some
brief remarks (Section VI).

\section{The Model}

The pyrochlore is a lattice of corner-sharing tetrahedra which can be constructed from the diamond lattice by placing a site at the midpoint of each bond (Fig. \ref{fig:geometry}). The result is a quadripartite structure, which may alternatively be described as an fcc lattice with a four-site basis. From the former construction, it is evident that the centers of the tetrahedra lie on the sites of the diamond lattice: in other words, the dual lattice of the pyrochlore is the diamond lattice -- a fact which we will make use of extensively below. We now turn to the Potts model
which we will introduce from the perspective of the Heisenberg model as this will yield a
vector spin representation of the Potts spins which will be central to this paper.

The pyrochlore lattice is highly frustrated from the perspective of classical collinear antiferromagnetism: the  nearest-neighbor classical Heisenberg antiferromagnet on this lattice has an extensive ground state degeneracy and remains a quantum
paramagnet at all temperatures \cite{MoessnerChalker}. Since they will be relevant to us, we briefly summarize some details of the Coulomb phase for $O(3)$ (Heisenberg) spins on the pyrochlore. The canonical nearest-neighbor Heisenberg Hamiltonian
\begin{equation}\label{eq:HHeisenberg}
H = J \sum_{\langle i, j \rangle}\mathbf{S}_i \cdot \mathbf{S}_j
\end{equation}
can be re-written, up to an overall constant, as
$$H= \frac{J}{2}\sum_{\boxtimes} \left( \sum_{i \in \boxtimes}\mathbf{S}_i\right)^{2},$$
where the sum in parenthesis runs over the four spins at the corners of each tetrahedron, and the outer sum runs over all tetrahedra in the lattice. Thus, the ground states are defined by spins satisfying \textit{local} constraints:
\begin{equation}
\sum_{i \in \boxtimes}S_{i}^{\alpha} = 0
\label{eq:constraint}
\end{equation}
 for each tetrahedron and each spin component $\alpha$.

 As a result of these local constraints, each ground state can be mapped to a configuration of divergence-free fluxes, one for each spin component, on the dual diamond lattice. Upon coarse-graining, the entropic cost of fluctuations within the ground state manifold leads to an emergent `electrodynamics' $-$ with the coarse-grained fluxes playing the role of divergence-free lattice electromagnetic fields. The process yields asymptotically dipolar forms for spin (and field) correlation functions, a hallmark of the celebrated ``Coulomb Phase''.

We now consider applying a symmetry breaking potential that restricts the Heisenberg spins to four symmetrically situated points in spin space (Fig. \ref{fig:spinspace}b). The spin on each pyrochlore site must now belong to the following set of four spins pointing towards the corners of a regular tetrahedron in spin-space:
\begin{eqnarray}
&\mathbf{S}_A = (-1,1,1); \quad
&\mathbf{S}_B = (1,-1,1);\nonumber\\
&\mathbf{S}_C = (-1,-1,-1); \quad
&\mathbf{S}_D = (1,1,-1).\label{eq:Pottsflavors}
\end{eqnarray}
Observe that any two (different) spins in (\ref{eq:Pottsflavors}) make an angle of $\cos^{-1}(-\frac{1}{3})$ with one another, so that the dot-product of any two spins in the set is $\mathbf{S}_\alpha\cdot\mathbf{S}_\beta = 4\,\delta_{\alpha\beta} -1$ where $\alpha, \beta = A, B, C,$ or $D$. Thus the nearest neighbor interaction energy has the character of an antiferromagnetic
Potts interaction between four states: it prefers neighbors to be different but is indifferent
to how that is achieved. Formally, the Hamiltonian (\ref{eq:HHeisenberg}) with the spins restricted to the set (\ref{eq:Pottsflavors}) is equivalent to the Hamiltonian
\begin{equation}\label{eq:HPotts}
H_P = J \sum_{\langle i, j\rangle} \delta_{\sigma_i, \sigma_j}.
\end{equation}
where the Potts spins $\sigma_i$ can be in one of four states: $A, B, C$ or $D$. In
essence we have mapped from Potts variables to a set of vector spins. The ground state constraint
(\ref{eq:constraint}) restricted to the set (\ref{eq:Pottsflavors}) is equivalent to the
condition that each tetrahedron to contain all four Potts states.

The Potts model has a discrete macroscopic ground state degeneracy---a remnant of the
continuous macroscopic degeneracy of the $O(3)$ model\cite{MoessnerChalker}. To show
this, a strict lower bound on the entropy can be obtained by using the degeneracy of the three-state Potts model on the kagome\cite{HuseRutenberg, Baxter} and by viewing the pyrochlore as alternating layers of kagome and triangular planes. The result is the bound\cite{ParameswaranAKLT} $\Omega > 4\,(1.208\, 72)^{N/2}$, corresponding to an entropy $S/ k_B N > (1/2)\log(1.208\, 72).$
A more direct estimate is the Pauling entropy\cite{Pauling} for this system. For a given tetrahedron, $4!$ of the possible $4^4$ states are ground states. Treating the tetrahedral constraints as independent gives a ground state degeneracy $$\Omega = 4^N \, \left(\frac{4!}{4^4}\right)^{N_{tet}} = \left(\frac{3}{2}\right)^{N/2},$$ where $N$ is the number of spins and $N_{tet} = N/2$ is the number of tetrahedra. This corresponds to
an entropy per spin $S/k_B N = (1/2)\log(3/2)$ which is, interestingly, the same as the Pauling estimate for the entropy of
spin ice \cite{SLSspinice}. 
As advertised in the introduction, the reduction from Heisenberg spins to Potts spins suggests
that the latter system will still exhibit a Coulomb phase. We now turn to a precise formulation
of this conjecture.

\section{Flux Fields and Correlations}
\subsection{Flux Fields}

Our development of  flux fields closely parallels  the construction in the case of the $O(N)$ antiferromagnet \cite{Isakov2004, Henley2005}. The essential idea is to map the spin variables to a flux field defined on the sites of the dual diamond lattice. The local ground state constraint (\ref{eq:constraint}) -- which still applies to the Potts spins as defined in (\ref{eq:Pottsflavors}) -- is then mapped into a requirement that the flux configuration be divergence-free.

\begin{figure}
 \includegraphics[width=7.5cm]{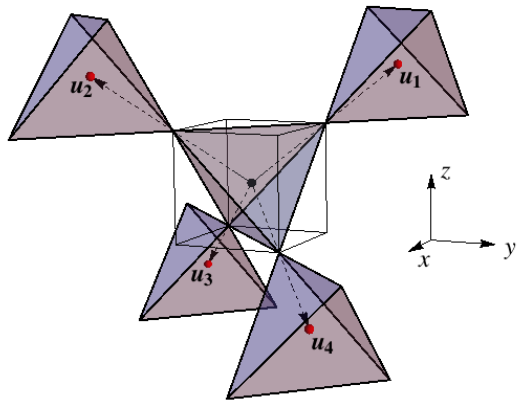}
 \caption{\label{fig:geometry}
 Geometry of the pyrochlore lattice. The centers of the up and down tetrahedra define the two sublattices of the diamond lattice, denoted by black and red dots. The pyrochlore lattice sites lie at the vertices of the tetrahedra. The vectors $\mathbf{u}_\kappa$ define the bond vectors of diamond.   }
 \end{figure}

We begin by defining bond vectors $\mathbf{u}_{\kappa}$ pointing from the even to the odd sublattices of the bipartite diamond lattice (i.e., from the centers of `up' to `down' tetrahedra on the pyrochlore), which take the form
\begin{align}
&\mathbf{u}_1 = \left(-\frac{1}{4},\frac{1}{4},\frac{1}{4}\right);
&\mathbf{u}_2 = \left(\frac{1}{4},-\frac{1}{4},\frac{1}{4}\right); \nonumber\\
&\mathbf{u}_3 = \left(-\frac{1}{4},-\frac{1}{4},-\frac{1}{4}\right);
&\mathbf{u}_4 = \left(\frac{1}{4},\frac{1}{4},-\frac{1}{4}\right)
\label{eq:uvecs}
\end{align}
 where the fcc lattice constant has been chosen as $a = 1$. The spins live on the midpoints of the bonds; Fig.~\ref{fig:geometry} illustrates the geometry of the lattice and the bond vectors.

Next,  we define three vector flux fields on each bond, one for each spin component of the Potts spins represented in (\ref{eq:Pottsflavors}):
 \begin{equation}
 \mathbf{B}_{\kappa}^{\alpha} = S_{\kappa}^{\alpha}\mathbf{u}_{\kappa},
 \label{eq:fluxDef}
 \end{equation}
 where $\mathbf{S}_{\kappa}$ denotes the spin on bond $\mathbf{u}_{\kappa}$, and $\alpha = 1,2,3$ labels the spin components. The flux field on a site of the diamond lattice is defined as the sum of the fields on the four tetrahedral bonds emerging from that site,
 \begin{equation}
 \mathbf{B}^{\alpha}(\mathbf{R}) = \sum\limits_{\kappa = 1}^{4} S_{\kappa}^{\alpha}(\mathbf{R})\mathbf{u}_{\kappa},
 \label{eq:fluxDefDiamond}
 \end{equation}
 where $\mathbf{R}$ is a diamond lattice vector, and $\kappa$ sums over the four tetrahedral sites (on pyrochlore) surrounding the diamond site (tetrahedron center). The mapping from spin to flux variables is invertible (see Appendix).

From our definition of the Potts spins, (\ref{eq:Pottsflavors}), we see that in any ground state, for each spin component $S^\alpha$, we have two ``incoming'' $(+1)$ and two ``outgoing'' $(-1)$ spins on each tetrahedron, i.e.  each tetrahedron obeys a two-in, two-out `ice rule' for each spin component.
It follows from this, and our definition of the flux fields, that the local Potts constraint maps to a zero-divergence condition for each of the $\mathbf{B}^{\alpha}$ fields, $\nabla\cdot \mathbf{B}^{\alpha} =0$. In an electrodynamic representation it is appropriate to refer to these flux fields as `magnetic' fields and then their sources will be monopoles---this is the
nomenclature that is natural in the context of spin ice and is the one we will use
here\footnote{Alternatively we could just as well refer to them as electric fields, sources
as electric charges and the constraint as Gauss's law. This is more natural in thinking of
quantum models where there is a natural identification with lattice gauge theories.}.

Naively, the problem just looks like three copies of spin ice, one for each spin component. Of course, the three components are not independent and therefore we might expect some correlation between the three magnetic fields. However, we will see shortly that our naive expectation is justified:  at long distances, these cross-correlations vanish and the physics is described by three independent divergence-free `Maxwell' fields.

\subsection{Coarse-grained Free Energy and Correlations}
Thus far, we have given a characterization of the ground state manifold in terms of divergence-free configurations of three magnetic fields. However, in order to compute correlations in the limit
$T\rightarrow 0$, we need a more workable description of the ground state manifold which we
will now obtain by coarse graining.

 Let us consider one of the three magnetic fields, say $\mathbf{B}^{1}$. If we were to flip the direction of flux on one of the ``in'' bonds at a diamond site (say by switching spins $S_A$ and $S_B$), the zero divergence condition would require us to also flip the direction of an ``out'' bond at the site. We can continue flipping spins in such fashion until we get either a closed loop of $S_A$, $S_B$ spins, or a string of $S_A$, $S_B$ spins that extends across the entire system (for finite systems, the latter eventually closes through periodic boundary conditions). Flipping spins that lie on closed loops leaves the net magnetic flux through the system unchanged, while flipping spins on spanning strings changes the net flux threading the system. Since the average flux contributed by a closed loop is zero, systems with large numbers of closed ``flippable'' loops will have a small  net $\mathbf{B}^{1}$. On the other hand, a large and saturated net $\mathbf{B}^{1}$ requires the field on almost every site of the dual lattice to point in the same direction and therefore the number of flippable loops is small -- an intuitive picture is that a the only flippable loops in a  saturated field configuration are those that span the system.

Thus far, we have only talked of lattice magnetic fields that live on the bonds of the pyrochlore. To derive long-wavelength properties, we need to define smoothly varying continuum fields. We do this by coarse-graining - the field $\mathbf{\tilde B}^{\alpha}(\mathbf{r})$ is defined as the average of the lattice fields $\mathbf{B}^{\alpha}$ in some neighborhood of $\mathbf{r}$ that is much larger than the lattice spacing but much smaller than the system size. The discrete constraint naturally translates into a divergence-free constraint for the coarse-grained fields.  

 Microscopically, there are many more configurations consistent with a small net coarse-grained $\mathbf{\tilde B}^{1}$ rather than a large saturated $\mathbf{\tilde B}^{1}$; the same arguments obviously apply to all three magnetic fields. Therefore configurations with small average fields are entropically favored. To lowest order, the (entirely entropic) free energy  as a function of the coarse-grained fields and consistent with symmetries can be written as
\begin{eqnarray}\label{eq:MaxwellFreeEnergy}
\lefteqn{F_{tot}({\mathbf{\tilde B}^{\alpha}(\mathbf{r})}) = - T S}  \nonumber \\
&=& \frac{1}{2}\frac{\kappa \,T}{a} \int d^3r \left({\left| \mathbf{\tilde B}^{1}(\mathbf{r})\right|}^2 + {\left| \mathbf{\tilde B}^{2}(\mathbf{r})\right|}^2 + {\left| \mathbf{\tilde B}^{3}(\mathbf{r})\right|}^2\right)
\end{eqnarray}
where we've inserted a factor of the lattice-spacing $a$ to make the stiffness, $\kappa$, dimensionless. Since we are restricting our attention for the moment to ground state configurations, the coarse-grained fields still satisfy the zero-divergence constraint, $\nabla \cdot  \mathbf{\tilde B}^{\alpha} =0$. The free energy (\ref{eq:MaxwellFreeEnergy})  coupled with the divergence-free constraint yields three copies of Maxwell electrodynamics in a standard fashion. Introducing three vector potentials $\mathbf{A}^\alpha$ to implement the constraints, we can rewrite the free energy as
\begin{equation*}
F =   \frac{1}{2} \frac{\kappa \, T}{a} \int d^3r \sum\limits_{\alpha = 1}^{3}     {\left| \nabla \times \mathbf{A}^\alpha(\mathbf{r})\right|}^2
\end{equation*}

We calculate the long distance correlators of the magnetic fields, and the result is the dipolar form typical for Coulomb phases
\begin{eqnarray}
G^{\alpha \beta}_{i j} (\mathbf{r}) &\equiv& \langle \tilde B^{\alpha}_{i}(\mathbf{r})\tilde B^{\beta}_{j}(\mathbf{0})\rangle \nonumber\\
 &=&\frac{a}{4\pi\kappa} \delta_{\alpha \beta} \, \frac{3r_i r_j - r^2 \delta_{ij}}{r^5}
\label{eq:correlation}
\end{eqnarray}
where $i,j$ refer to the $x,y,z$ components of each magnetic-field.

\begin{figure}
 \includegraphics[width=\columnwidth]{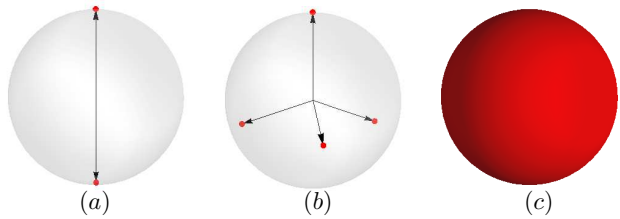}
 \caption{\label{fig:spinspace} Spin space representation of Ising, Potts and Heisenberg spins (left to right). The red areas represent the accessible portions of spin space - Ising spins are confined to lie on only two points, 4-state Potts spins occupy only four points, and Heisenberg spins can lie anywhere on the sphere. While Ising spins only require a single gauge field for their description, Heisenberg spins require three such fields. Somewhat surprisingly  Potts spins -- which {\it a priori} would appear to have a much reduced symmetry compared to the Heisenberg case -- also require three gauge fields.} 
  \end{figure}

Finally, we reiterate that Ising spins, which occupy only two collinear points on a sphere in spin-space, require a single magnetic field $\mathbf{B}$ to describe their Coulomb phase. Classical $O(3)$ spins can lie anywhere on a sphere, and they require three fields $\mathbf{B}^1$, $\mathbf{B}^2$ and $\mathbf{B}^3$ for their description -- one for each spin component. It is interesting that even though Potts spins are locked to just 4 points in spin space, they still require three magnetic fields for their complete description: the theory thus renormalizes at low temperatures to an effective $O(3)$ symmetry. Fig.~\ref{fig:spinspace} illustrates this idea.

\subsection{Monte Carlo Simulations} 
In order to test our conjectured form (\ref{eq:correlation}) for the real-space correlation functions for the magnetic field, we turn now to a Monte Carlo study of the ground state manifold. We perform simulations on systems with   $L\times L\times L$ unit cells with $ L = 8, 16, 32$ and periodic boundary conditions; since there are 4 sites per pyrochlore unit cell, this corresponds to  $4 L^3$ spins. The simulations were performed using a standard ``worm-update" algorithm.
In each update step we first identify a closed ``worm'' of alternating spin flavors (for instance $AB\ldots AB$), and then flip all the spins along the worm (i.e. interchange the spins $A\leftrightarrow B$). This move respects the Potts constraint since each bond in the worm is only part of a single tetrahedron, and exchanging spin-flavors on a bond still leaves a tetrahedron with all four Potts flavors. There are ${4\choose 2} = 6$ types of worms, and the starting site for a worm and its type were chosen randomly for each update. It is
instructive to think of this in the language of fluxes: interchanging $A$ and $B$ sites corresponds, via
Eq.~(\ref{eq:Pottsflavors}), to identifying a closed loop of type 1,2 and 3 fluxes and then reversing
the first two of these. Reversing closed loops of fluxes clearly leaves the solenoidal constraints intact.

We simulate $M$ independent Markov chains, each with $N$ configurations along the chain ($M$ and $N$ were typically 100 and 10,000 respectively). To generate these, we begin with $M$ independent ``seed'' ground state configurations and use
the worm update described above
 to generate the states along the chain. To ensure that successive states along the chain were roughly independent, we perform several ($\sim 30$) such worm updates before recording a new configuration which is then added to the chain.

In analyzing the data,  we first obtain the average value of the correlation functions for each Markov chain, and then average these across all $M$ chains. The error  is estimated as the standard error of the single-chain averaged correlation function across the $M$ independent chains.

We compute the average correlation function $G^{\alpha \beta}_{i j} (\mathbf{r})= \langle  B^{\alpha}_{i}(\mathbf{x+ r}) B^{\beta}_{j}(\mathbf{x})\rangle$ in two independent directions $\mathbf{r}$ for all 36 combinations of $\alpha, \beta, i,j$. The vectors $\mathbf{r}$ were chosen as $\mathbf{r}_1=n_1 \left(0, \frac{1}{2}, \frac{1}{2}\right)$ and $\mathbf{r}_2=n_2 \left(0,0,1\right)$ with $n_1, n_2 \in {Z}$.  These correspond to the fcc lattice vectors $\mathbf{r}_1 = n_1 \mathbf{a}_1$ and $\mathbf{r}_2 = n_2 (\mathbf{a}_1+ \mathbf{a}_2- \mathbf{a}_3)$ where the $\mathbf{a}_i$ are fcc basis vectors with lattice constant $a=1$.

\begin{figure}
 \includegraphics[width=\columnwidth]{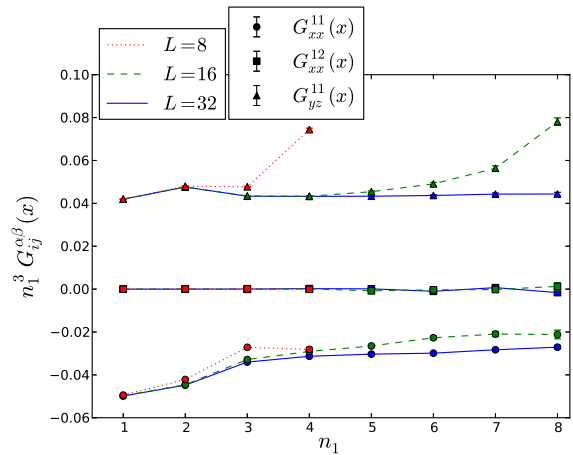}
 \caption{\label{fig:corr1} Monte Carlo data for three representative correlation functions $G^{11}_{xx}(\mathbf{r_1}), G^{12}_{xx}(\mathbf{r_1})$ and $G^{11}_{yz}(\mathbf{r_1})$ in the direction $\mathbf{r}_1=n_1 \left(0, \frac{1}{2}, \frac{1}{2}\right)$ for lattice sizes $L = 8, 16, 32$. The correlators are multiplied by the cube of the distance $n_1^3$, and the horizontal trends are consistent with the expected dipolar form (\ref{eq:correlation}). All cross-correlators $G^{\alpha\beta}_{ij}$ for $\alpha \neq \beta$ vanish (only $G^{12}_{xx}$ displayed) confirming the diagonal form for the free energy (\ref{eq:MaxwellFreeEnergy}).}
 \end{figure}

\begin{figure}
 \includegraphics[width=\columnwidth]{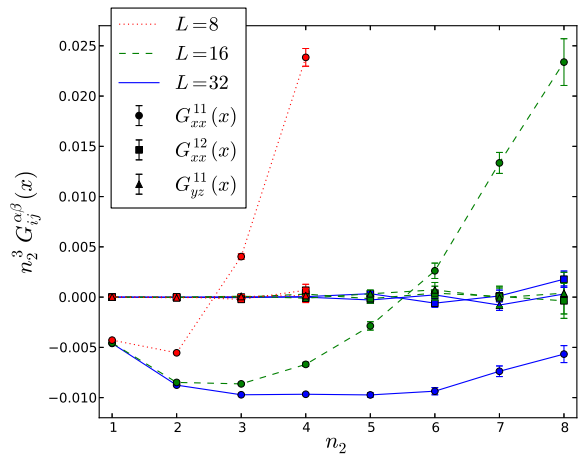}
\caption{\label{fig:corr2} Monte Carlo data for three representative correlation functions $G^{11}_{xx}(\mathbf{r_2}), G^{12}_{xx}(\mathbf{r_2})$ and $G^{11}_{yz}(\mathbf{r_2})$ in the direction $\mathbf{r}_2=n_2 \left(0,0,1\right)$ for lattice sizes $L = 8, 16, 32$. Once again, the correlators are multiplied by the cube of the distance $n_2^3$, and the trends are consistent with the expected dipolar form (\ref{eq:correlation}). Note that the correlations are weaker in the direction $\mathbf{r}_2$ as compared to $\mathbf{r}_1$ because same values of $n$ for the two cases correspond to larger physical distances $r_2$.}
 \end{figure}

The correlations fall off as $1/r^3$ consistent with the dipolar form (\ref{eq:correlation}). Figs. \ref{fig:corr1} and \ref{fig:corr2} show the representative correlators $G^{11}_{xx} (\mathbf{r})$, $G^{11}_{xy} (\mathbf{r})$, and $G^{12}_{xx} (\mathbf{r})$ multiplied by $n_i^3$, in the two directions $\mathbf{r}_1$ and $\mathbf{r}_2$. The agreement with the dipolar form is best in the regime $a \ll r \ll L$. The cross-correlators $G^{\alpha \beta}_{i j} (\mathbf{r})$ for $\alpha \neq \beta$ vanish in all directions for all $i,j$, confirming the diagonal form for the effective free energy (\ref{eq:MaxwellFreeEnergy}). We also checked that the ratios of correlations for different $i,j$ asymptote to the values predicted by (\ref{eq:correlation}). For example, $G^{11}_{xx}/G^{11}_{zz} \rightarrow -2$ in the direction $\mathbf{r}_1$.

Finally, we use correlation data to numerically estimate a value for the stiffness, $\kappa/a$ through (\ref{eq:correlation}). Correlations in the direction $\mathbf{r}_1$ yielded an average stiffness $\kappa_1 = 7.38 \pm 0.46$, while correlations in the direction $\mathbf{r}_2$ gave $\kappa_2 = 8.13\pm 0.93$; the values in the two directions are equal within the margins of error. The lattice constant $a$ is set to unity.

\section{Charges, defects and Dirac Strings}
While the effective free energies for the Heisenberg and Potts models are the same, the two differ in the nature of their excitations. Excitations above the ground state Potts manifold are gapped and quantized, with ``bionic'' defects that are charged under two of the three gauge fields.

In the ground states, each site of the dual diamond lattice has two incoming and two outgoing fluxes for each of the three magnetic fields. We can create defects by violating the zero-divergence constraint at a dilute set of points in the lattice. Such defects are ``charged'' under the different fields, with positive (negative) charges equal to the net outgoing (incoming) fluxes of each type at the defect. This is the usual charge in the sense of Gauss's law: each such charge represents a source of magnetic flux, and a violation of the divergence-free constraint for at least one of the fields.

It is convenient to first catalog defects in the Potts language: a defect arises when the four spins surrounding a dual lattice site violate the Potts rule. The simplest defects are those in which one spin flavor is repeated on a tetrahedron; for instance, we can have a defect tetrahedron with spins $S_B, S_B, S_C$ and $S_D$. There are twelve such defect tetrahedrons: we have four choices for the spin flavor that gets repeated ($S_B$ in our example), and three choices for the flavor that the repeated spin replaces ($S_A$ in the example).

\begin{figure}
 \includegraphics[width=0.75\columnwidth]{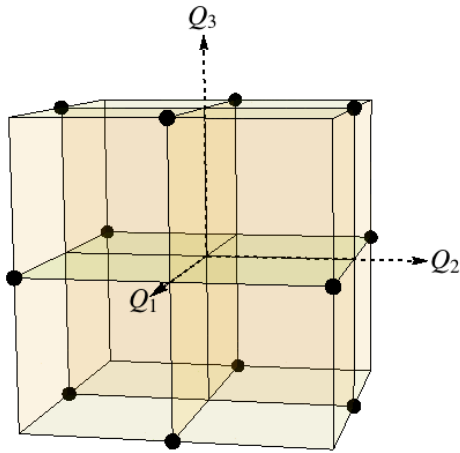}
 \caption{\label{fig:charges} Charges of the twelve types of bions (black dots) shown in $Q_1, Q_2, Q_3$ space, where $Q_i$ represents the charge under the field $B^i$. Each bion is charged under two gauge fields. The twelve charges naturally map to six oriented edges of a tetrahedron.}
 \end{figure}

Each of these defects have different charges under the three gauge fields. Looking at the spins in (\ref{eq:Pottsflavors}), we see that an ``up'' tetrahedron with $S_B, S_B, S_C, S_D$ has spin components $S^x = (1,1,-1,1)$, $S^y = (-1,-1,-1,1)$ and $S^z = (1,1,-1,-1)$. Thus, its charges under the three magnetic fields $\mathbf{B}^1$, $\mathbf{B}^2$ and $\mathbf{B}^3$ are $Q_1 = +2$, $Q_2 = -2$ and $Q_3 = 0$ respectively.  All twelve defects have a similar structure, in that they are doubly charged under two of the three magnetic fields -- hence the name bions. Table \ref{table:charges} catalogues the charges of the different bions. (These charges are reversed for corresponding defects on ``down'' tetrahedra, since the sense of ``in'' and ``out'' flux is reversed). Figure~\ref{fig:charges} depicts the 12 possible bions in $Q_1, Q_2, Q_3$ space.

\begin{figure*}
 \centering
 \includegraphics[width=2\columnwidth]{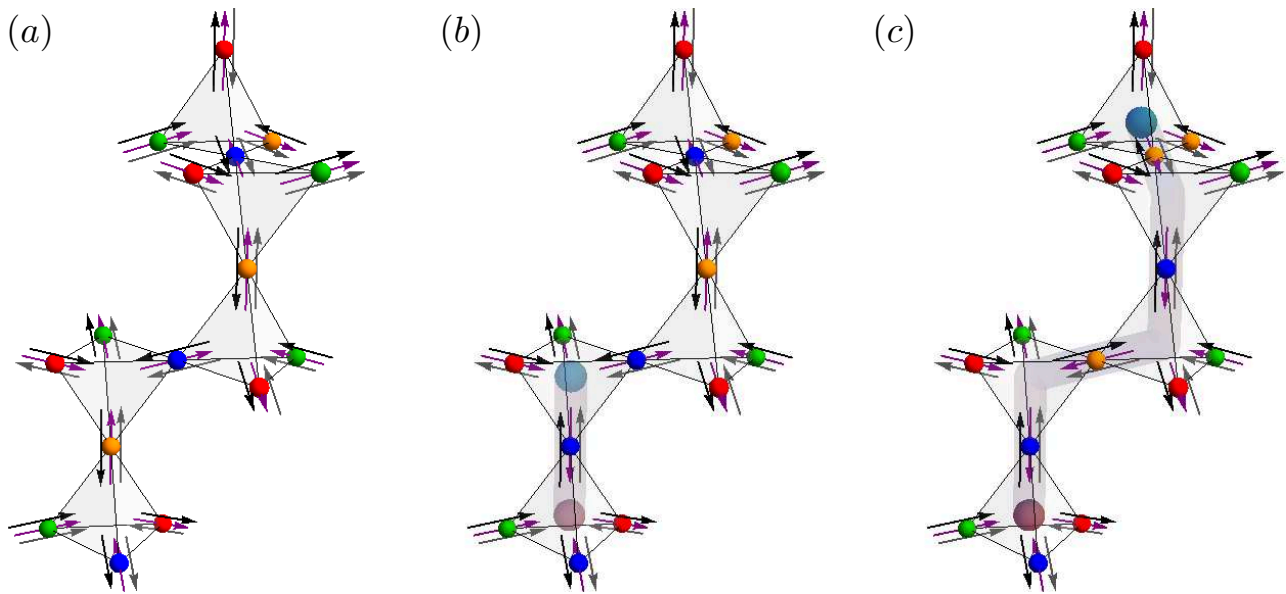}
 \caption{\label{fig:DiracString} Dirac string connecting two bions. The colors orange
 , blue, green and red represent the spin flavors $S_A, S_B, S_C$ and $S_D$ respectively. The three arrows on each site represent the three magnetic fields $B^1$ (black), $B^2$ (purple), and $B^3$ (gray). (a) Ground state configuration of the 4-state Potts model with no defects. Each tetrahedron has all four spin flavors, and all three magnetic fields obey two-in-two-out rules at each tetrahedron. (b) Switching $S_A$ (orange) to $S_B$ (blue) on the bottom tetrahedron creates (BBCD) type bions on two adjoining tetrahedra (red and blue spheres). The magnetic fields are no longer divergence-free, and the bion on the up tetrahedron has charges $Q_1 = +2$, $Q_2 = -2$, $Q_3 = 0$ (opposite charges for the bion on the down tetrahedron). (c) The bions move apart by flipping a trail of $S_A$, $S_B$ spins. The Dirac string acts as a flux tube carrying $B^1$ and $B^2$ type magnetic fluxes between the two bions.}
 \end{figure*}

Charge conservation demands that the defects are always created in oppositely charged pairs. One way to do this is to imagine creating a pair of bions by exchanging spins along a ``Dirac string'' containing two flavors of spin. For instance, we create a BBCD defect by replacing an $A$ spin with a $B$ spin on an ``up'' tetrahedron. This creates a second, oppositely charged defect on the adjoining ``down'' tetrahedron. The second defect can be moved away from the first by continually flipping a string of $A$, $B$ type spins. The second defect is of type AACD when it lies on another ``up'' tetrahedron; we may verify that the tetrahedron AACD carries opposite charge $Q_1 = -2$, $Q_2 = +2$ and $Q_3 = 0$. We can think of the Dirac string as a flux tube carrying two flavors of flux ($\mathbf{B}^1$ and $\mathbf{B}^2$ in our example) that connects bions that are oppositely charged under two magnetic fields. Figure~\ref{fig:DiracString} shows a Dirac string connecting two bions.

\begin{table}[h]
\centering
\begin{tabular}{c l p{5cm}}
\hline \hline
Type of Defect & \multicolumn{2}{c}{Charges and Flux Lines}  \\ [0.5ex]
\hline
 & $Q_1 = -2$ & \rdelim\}{6}{2cm}{\begin{center}Connected by a Dirac string of type A D\end{center} }\\[-11ex]
 A A B C & $Q_2 = 0$ &\\[-1ex]
 & $Q_3 = +2$ & \\[0.5ex]
 & $Q_1 = +2$ & \\[-1ex]
 D D B C & $Q_2 = 0$ &\\[-1ex]
 & $Q_3 = -2$ & \\[2ex]

  & $Q_1 = 0$ & \rdelim\}{6}{2cm}{\begin{center}Connected by a Dirac string of type A C\end{center} }\\[-11ex]
  A A B D & $Q_2 = +2$ &\\[-1ex]
  & $Q_3 = +2$ & \\[0.5ex]
  & $Q_1 = 0$ & \\[-1ex]
  C C B D & $Q_2 = -2$ &\\[-1ex]
  & $Q_3 = -2$ & \\[2ex]

  & $Q_1 = -2$ & \rdelim\}{6}{2cm}{\begin{center}Connected by a Dirac string of type A B\end{center} }\\[-11ex]
  A A C D & $Q_2 = +2$ &\\[-1ex]
  & $Q_3 = 0$ & \\[0.5ex]
  & $Q_1 = +2$ & \\[-1ex]
  B B C D & $Q_2 = -2$ &\\[-1ex]
  & $Q_3 = 0$ & \\[2ex]

  & $Q_1 =0$ & \rdelim\}{6}{2cm}{\begin{center}Connected by a Dirac string of type B D\end{center} }\\[-11ex]
  B B A C & $Q_2 = -2$ &\\[-1ex]
  & $Q_3 = +2$ & \\[0.5ex]
  & $Q_1 =0$ & \\[-1ex]
  D D A C & $Q_2 = +2$ &\\[-1ex]
  & $Q_3 = -2$ & \\[2ex]

  & $Q_1 =+2$ & \rdelim\}{6}{2cm}{\begin{center}Connected by a Dirac string of type B C\end{center} }\\[-11ex]
  B B A D & $Q_2 = 0$ &\\[-1ex]
  & $Q_3 = +2$ & \\[0.5ex]
  & $Q_1 =-2$ & \\[-1ex]
  C C A D & $Q_2 = 0$ &\\[-1ex]
  & $Q_3 = -2$ & \\[2ex]

  & $Q_1 =-2$ & \rdelim\}{6}{2cm}{\begin{center}Connected by a Dirac string of type C D\end{center} }\\[-11ex]
  C C A B & $Q_2 = -2$ &\\[-1ex]
  & $Q_3 = 0$ & \\[0.5ex]
  & $Q_1 =+2$ & \\[-1ex]
  D D A B & $Q_2 = +2$ &\\[-1ex]
  & $Q_3 = 0$ & \\[2ex]
  \hline
\end{tabular}
\caption{Catalog of defects and charges. Each defect is charged under two gauge fields - hence they are called bions.}
\label{table:charges}
\end{table}

Additionally, we can also imagine creating composite defects by adding two or more of the twelve fundamental bions. The composite charges form an fcc lattice which is a natural extension of Figure \ref{fig:charges}. There is an intuitive geometric picture for understanding the charge structure of the bions. The four spins point to the four corners of a tetrahedron in spin-space (and all spin-components sum to zero); replacing $S_A$ with $S_B$ to create a defect gives a vector of charges $\mathbf{Q}= -\mathbf{S}_A +\mathbf{S}_B$ under the different gauge fields. The charge $\mathbf{Q}$ thus corresponds to an edge of the spin-space tetrahedron. In this way, all twelve fundamental bions can be mapped to six \textit{oriented} edges of a tetrahedron in spin-space.

The bions act as sources of magnetic flux and experience a Coulomb force under each of the three magnetic fields. The origin of this force is purely entropic in nature. We imagine a finite number $N_b$ of bions scattered throughout the lattice at positions $\mathbf{r}_1, \mathbf{r}_2, \cdots \mathbf{r}_{N_b}$. We can compute the partition function $Z$ by integrating over all configurations of magnetic fields consistent with the distribution of bions. This gives the free energy of the sources in accordance with $Z/Z_0= e^{-F_{int}/T}$, where $Z_0$ is the partition-function in the absence of bions. Explicitly this yields:
\begin{eqnarray}
F_{int} &=& \sum\limits_{\substack{i=1\\j < i}}^{N_b} \frac{a \, T}{4 \pi\kappa} \frac{\left(Q_1^i Q_1^j+ Q_2^i Q_2^j+ Q_3^i Q_3^j\right)}{|\mathbf{r}_i- \mathbf{r}_j |} \nonumber\\
&+&  \sum\limits_{i=1}^{N_b} \alpha \,\frac{a\,T}{4 \pi\kappa} \; \frac{ (Q^i)^2}{a}
\label{eq:interaction}
\end{eqnarray}
where $i,j$ sum over all pairs of bions and $Q_\alpha^i$ represents the charge of the $i$th bion under the field $\mathbf{B}^\alpha$. The first term in (\ref{eq:interaction}) represents the Coulomb interaction energy of pairs of bions separated by distance $|\mathbf{r}_i- \mathbf{r}_j|,$ while the second term is the free energy of individual, isolated bions arising from self-interaction in the
field theory. The self-interaction term has an ultraviolet ambiguity, represented by the unknown
constant $\alpha$, which (naturally) cannot be fixed by the coarse-grained free energy alone.

Also note that (\ref{eq:interaction}) predicts that the interaction energy between a pair of bions is sensitive to their type.  For example, bions of type $AABC$ and $DDBC$ (equally and oppositely charged under $\mathbf{B}_1$ and $\mathbf{B}_3$) interact more strongly than say $AABC$ and $AABD$ which are charged under different gauge fields.

This is a good place to briefly comment on the finite-temperature properties of the Potts spins - in particular the form of the correlation function Eq.~(\ref{eq:correlation}). We know that the dipolar form of the correlation function is derived from the divergence-free constraint on the magnetic fields. This constraint is exactly satisfied at $T=0$ and gradually weakened as the temperature $T$ is increased. Heuristically, we might expect the  correlation to be dipolar up to some (temperature dependent) correlation length $\xi(T)$, and decay exponentially on length scales longer than $\xi$. It is easily seen \cite{Isakov2005} that for Ising spins, the creation energy of gapped ice-rule violating defects (monopoles) yields $\xi \sim e^{2J/3T}$. On the other hand, for Heisenberg spins the gapless excitations yield much softer violations of the divergence-free rule and, correspondingly, a much shorter finite-temperature correlation length given by \cite{Gregor} $\xi \sim 1/\sqrt{T}$. For gapped Potts spins, the creation energy of a Bionic defect is 4J which gives $\xi(T) \sim e^{4J/3T}$; the gap to excitations helps preserve the dipolar form of the correlations to higher temperatures. As is typical, the gap also manifests itself in the exponential low-temperature decay of various thermodynamic quantities but these are not the focus of this paper.  

\section{Worm length distributions}

Coulomb phases come with a natural incipient loop structure---absent defects one can define closed lines
of flux thanks to the underlying conservation laws. This makes the statistics of the loops worthy of interest.
Indeed Jaubert, Haque and Moessner (JHM) \cite{Jaubert2011} have studied loop statistics for the ground
state manifold of spin ice and
found a characteristic scaling of their probability distribution which they have related to the properties of
random walks in three
dimensions. This suggests that this scaling might be more generally associated with Coulomb phases and we will investigate and verify that possibility here.

Specifically, JHM have studied the distribution of worm lengths for spin ice which is easily
done by keeping track of the worms used to update configurations in the Monte Carlo.
As in our problem, worms in spin ice are closed strings of alternating spin flavors but now with the %additional
feature that while there is only one species of worm, at each step there is a binary choice that must be made randomly\footnote{Ref.~\onlinecite{Jaubert2011} also studied closed loops of up and down spins alone but they
have no analog in the 4-state Potts problem.}. For these JHM found a characteristic
scaling for probability distribution of worm lengths $p(\ell)$,
\begin{equation}\label{loopdistt}
p(\ell) = \frac{1}{L^3}\; f\left(\frac{\ell}{L^2}\right)
\end{equation}
where $\ell$ refers to the worm length and $L$ is the linear dimension of the system size. This scaling form
unifies two populations of worms: short worms whose probability scales as $p(\ell) \sim \ell^{-3/2}$, and long
winding worms (that close after spanning the system through periodic boundary conditions) for which
$p(\ell) \sim L^{-3}$.

\begin{figure}
 \includegraphics[width=8cm]{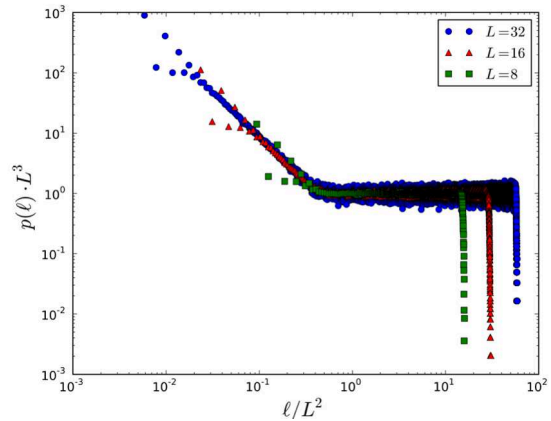}
 \caption{\label{fig:looplengths} Probability distribution of worm-lengths $\ell$ for system-sizes $L=8,16,32$. We clearly see two regimes -- an $\ell^{-3/2}$ scaling for short loops, and an $\ell$ independent scaling for winding loops.}
 \end{figure}

We have investigated analogous worm-length distributions in the Potts model. It should be noted that the Potts model has six species of worms (each worm has only two Potts spin flavors and ${4 \choose 2} = 6$). However, by symmetry, all six worm types have identical distributions in the Coulomb phase. Fig.~\ref{fig:looplengths} shows the worm-length distributions obtained for system sizes $L=8, 16,$ and $32$ plotted in scaled variables.
Leaving aside the deviations at very small and very large loops sizes we see that that the loop distribution
indeed obeys the scaling form (\ref{loopdistt}) which is then independent evidence that the Potts model
is in a Coulomb phase. We direct the reader to Ref.~\onlinecite{Jaubert2011} for
a rationalization of this scaling in terms of the properties of random walks.

\section{Concluding Remarks}
To summarize, we have shown that the classical antiferromagnetic four-state Potts model on the pyrochlore lattice is in a Coulomb phase described by three emergent gauge fields as $T\rightarrow 0$. It is instructive to view the Potts model as arising from a symmetry breaking potential that restricts $O(3)$ spins to just four points in spin space. Nevertheless, we have shown that the long-wavelength effective free energies of the $O(3)$ and Potts models are identical.

An important point of difference between the Heisenberg and Potts models lies in the nature of excitations above the ground state manifold. While the Heisenberg model involves gapless excitations, the Potts model has gapped excitations with a novel charge structure. We find 
twelve types of ``bionic'' defects, each charged under two of the three gauge fields. The charges are deconfined and can be connected by ``worm-like'' flux tubes of alternating spin flavor. We computed probability distributions for lengths of closed worms, and found scaling laws in accordance with previous diagnostics of the Coulomb phase.

The evident next step in this program is to incorporate quantum dynamics in a quantum Potts model exhibiting a Rokhsar-Kivelson point. We expect to report the details of this construction and
a study of the excitations and phase diagram in a future publication \cite{kmps-wip}.

\noindent {\it Note added.--} Upon conclusion of this work, we became aware of overlapping results of a study of the same problem\cite{chernwu-unpub}.
\begin{acknowledgments}We thank Bryan Clark and Michael Kolodrubetz for useful discussions on Monte Carlo simulations. SAP and SLS are grateful to Daniel Arovas for collaboration on related work. We acknowledge support from the Simons Foundation through the Simons Postdoctoral Fellowship at UC Berkeley (SAP) and from the National Science Foundation through grant number DMR-1006608 (SLS).
\end{acknowledgments}

\begin{appendix}

\section{Spin-Spin Correlation Functions}
It is useful to have a reference for converting the flux-field correlation functions (\ref{eq:correlation}) to spin-spin correlation functions for the Potts spins living on the four sublattices of pyrochlore. We have three flux fields $\mathbf{B}^1$, $\mathbf{B}^2$, $\mathbf{B}^3$ each with three components, for a total of nine flux components; these are labeled $B^{\alpha}_i$ for $\alpha = 1,2,3$ and $i = x, y,z$ in (\ref{eq:correlation}). One the other hand we have four Potts spins on every diamond site, each with three components, for a total of twelve spin components; we label these $S^{\alpha}_{\kappa}$ for $\alpha = 1,2,3$ and $\kappa = 1,2,3,4$. However, the constraint equations on the Potts spins (\ref{eq:constraint}) ensure that there are only nine \textit{independent} spin components, thereby permitting an invertible mapping from flux-fields to spins.

As an example, we use the definitions of the flux fields on the diamond sites (\ref{eq:fluxDefDiamond}) and the definitions of the bond vectors $
\mathbf{u}_\kappa$ (\ref{eq:uvecs}), to explicitly write the $x,y,z$ components of
$$
\mathbf{B}^1(\mathbf{r}) = S_1^x(\mathbf{r})\mathbf{u}_1 + S_2^x(\mathbf{r})\mathbf{u}_2+S_3^x(\mathbf{r})\mathbf{u}_3+S_4^x(\mathbf{r})\mathbf{u}_4\
$$
as

\begin{eqnarray}
B^1_x& = &\frac{1}{4}\left(-S_{1}^{x} + S_{2}^{x}-S_{3}^{x}+S_{4}^{x}\right) \nonumber\\
B^1_y &=& \frac{1}{4}\left(S_{1}^{x} - S_{2}^{x}-S_{3}^{x}+S_{4}^{x}\right) \nonumber\\
B^1_z &=& \frac{1}{4}\left(S_{1}^{x} + S_{2}^{x}-S_{3}^{x}-S_{4}^{x}\right) \nonumber
\end{eqnarray}
where we have dropped the explicit dependence on $\mathbf{r}$ to simplify our notation. Finally we impose the spin constraint equations (\ref{eq:constraint}) to eliminate $S_4^\alpha$ and we get
\begin{eqnarray}
B^1_x& = &\frac{1}{2}\left(-S_{1}^{x} -S_{3}^{x}\right) \nonumber\\
B^1_y &=& \frac{1}{2}\left( - S_{2}^{x}-S_{3}^{x}\right) \nonumber\\
B^1_z &=& \frac{1}{2}\left(S_{1}^{x} + S_{2}^{x}\right) \nonumber.
\end{eqnarray}
A similar exercise can be carried out for the $\mathbf{B}^2$ and $\mathbf{B}^3$ spins to give the following matrix of transformations:
\begin{widetext}
$$
\makebox[\columnwidth]{$
\begin{pmatrix}
B^1_x\\
B^1_y\\
B^1_z\\
B^2_x\\
B^2_y\\
B^2_z\\
B^3_x\\
B^3_y\\
B^3_z
\end{pmatrix} =
\left( \begin{array} {rrrrrrrrr}
-\frac{1}{2}&0&0&0&0&0&-\frac{1}{2}&0&0\\
0&0&0&-\frac{1}{2}&0&0&-\frac{1}{2}&0&0\\
\frac{1}{2}&0&0&\frac{1}{2}&0&0&0&0&0\\
0&-\frac{1}{2}&0&0&0&0&0&-\frac{1}{2}&0\\
0&0&0&0&-\frac{1}{2}&0&0&-\frac{1}{2}&0\\
0&\frac{1}{2}&0&0&\frac{1}{2}&0&0&0&0\\
0&0&-\frac{1}{2}&0&0&0&0&0&-\frac{1}{2}\\
0&0&0&0&0&-\frac{1}{2}&0&0&-\frac{1}{2}\\
0&0&\frac{1}{2}&0&0&\frac{1}{2}&0&0&0
\end{array}\right)  \begin{pmatrix}
S^x_1\\
S^y_1\\
S^z_1\\
S^x_2\\
S^y_2\\
S^z_2\\
S^x_3\\
S^y_3\\
S^z_3
\end{pmatrix}
$}
$$
We can invert the relation above to obtain the spins in terms of the flux fields as follows:
$$
\makebox[\columnwidth]{$
\begin{pmatrix}
S^x_1\\
S^y_1\\
S^z_1\\
S^x_2\\
S^y_2\\
S^z_2\\
S^x_3\\
S^y_3\\
S^z_3\\
\end{pmatrix} =
\left( \begin{array} {rrrrrrrrr}
-1&1&1&0&0&0&0&0&0\\
0&0&0&-1&1&1&0&0&0\\
0&0&0&0&0&0&-1&1&1\\
1&-1&1&0&0&0&0&0&0\\
0&0&0&1&-1&1&0&0&0\\
0&0&0&0&0&0&1&-1&1\\
-1&-1&-1&0&0&0&0&0&0\\
0&0&0&-1&-1&-1&0&0&0\\
0&0&0&0&0&0&-1&-1&-1\\
\end{array} \right)
\begin{pmatrix}
B^1_x\\
B^1_y\\
B^1_z\\
B^2_x\\
B^2_y\\
B^2_z\\
B^3_x\\
B^3_y\\
B^3_z
\end{pmatrix}
$}
$$\end{widetext}
These relations allow us to express spin-spin correlation functions as simple linear combinations of the ${B}^\alpha_i$ correlations.

\section{Symmetries and Stability of the Free Energy}
We have claimed in the bulk of the paper that the long-wavelength coarse-grained free energy is insensitive to breaking Heisenberg ($O(3)$) symmetry down to a Potts symmetry. In this section, we formalize this claim using a renormalization-group argument. We will show that the lowest order terms that can be added to the free energy (and are allowed by symmetry considerations) are irrelevant in an RG sense. 

Consider a given ground state configuration of Potts spins on the pyrochlore lattice. Permuting the spins (say by exchanging spins of type $S_A$ and $S_B$ on every tetrahedron) gives another ground state configuration. Each such ``internal'' symmetry of the microscopic spin degrees of freedom should translate into a symmetry of the coarse-grained $\mathbf{\tilde B}$ fields\footnote{We choose a coarse-graining procedure that respects the microscopic symmetries.} in accordance with Appendix A. In fact, six permutation group elements are required to generate all $4! = 24$ permutations of the Potts spins; these map to the following symmetries of the free energy: 
\begin{eqnarray}
f(\mathbf{\tilde B}_1, \mathbf{\tilde B}_2, \mathbf{\tilde B}_3) &=& f(\mathbf{\tilde B}_2, \mathbf{\tilde B}_1, \mathbf{\tilde B}_3) \nonumber\\ 
&=& f(\mathbf{\tilde B}_3, \mathbf{\tilde B}_2, \mathbf{\tilde B}_1) \nonumber\\
&=& f(\mathbf{\tilde B}_1, \mathbf{\tilde B}_3, \mathbf{\tilde B}_2) \nonumber\\
&=& f(-\mathbf{\tilde B}_1, -\mathbf{\tilde B}_2, \mathbf{\tilde B}_3) \nonumber\\
&=& f(-\mathbf{\tilde B}_1, \mathbf{\tilde B}_2, -\mathbf{\tilde B}_3) \nonumber\\
&=& f(\mathbf{\tilde B}_1, -\mathbf{\tilde B}_2, -\mathbf{\tilde B}_3)
\label{eq:sym1}
\end{eqnarray} 

An example best illustrates how one arrives at the symmetries in Eq.~(\ref{eq:sym1}). A given tetrahedron has magnetic fields defined by 
\begin{eqnarray*}
\mathbf{B}_1 &=& S_A^x \; \mathbf{u}_{a} + S_B^x \; \mathbf{u}_{b}+ S_C^x \; \mathbf{u}_{c}+ S_D^x \; \mathbf{u}_{d}\\
&=&  -\mathbf{u}_{a} + \mathbf{u}_{b}- \mathbf{u}_{c}+ \mathbf{u}_{d} 
\end{eqnarray*}
\begin{eqnarray*}
\mathbf{B}_2 &=& S_A^y \; \mathbf{u}_{a} + S_B^y \; \mathbf{u}_{b}+ S_C^y \; \mathbf{u}_{c}+ S_D^y \; \mathbf{u}_{d}\\
&=&  +\mathbf{u}_{a} - \mathbf{u}_{b}- \mathbf{u}_{c} + \mathbf{u}_{d}
\end{eqnarray*}
\begin{eqnarray*}
\mathbf{B}_3 &=& S_A^z \; \mathbf{u}_{a} + S_B^z \; \mathbf{u}_{b}+ S_C^z \; \mathbf{u}_{c}+ S_D^z \; \mathbf{u}_{d}\\
&=&  +\mathbf{u}_{a} + \mathbf{u}_{b}- \mathbf{u}_{c} - \mathbf{u}_{d}
\end{eqnarray*}
where $\{a, b, c, d\} \in$ Permutation$\{1,2,3,4\}$ label the $\mathbf{u}$ sublattice vectors on which the spins $\{S_A, S_B, S_C, S_D\}$ live. Then, exchanging spins $S_A$ and $S_B$ leads to the modified fields

\begin{eqnarray*}
\mathbf{B}'_1 &=& S_B^x \; \mathbf{u}_{a} + S_A^x \; \mathbf{u}_{b}+ S_C^x \; \mathbf{u}_{c}+ S_D^x \; \mathbf{u}_{d}\\
&=&  +\mathbf{u}_{a} - \mathbf{u}_{b}- \mathbf{u}_{c}+ \mathbf{u}_{d} \\ &=& \mathbf{B}_2
\end{eqnarray*}
\begin{eqnarray*}
\mathbf{B}'_2 &=& S_B^y \; \mathbf{u}_{a} + S_A^y \; \mathbf{u}_{b}+ S_C^y \; \mathbf{u}_{c}+ S_D^y \; \mathbf{u}_{d}\\
&=&  -\mathbf{u}_{a} + \mathbf{u}_{b}- \mathbf{u}_{c} + \mathbf{u}_{d} \\&=& \mathbf{B}_1
\end{eqnarray*}
\begin{eqnarray*}
\mathbf{B}'_3 &=& S_B^z \; \mathbf{u}_{a} + S_A^z \; \mathbf{u}_{b}+ S_C^z \; \mathbf{u}_{c}+ S_D^z \; \mathbf{u}_{d}\\
&=&  +\mathbf{u}_{a} + \mathbf{u}_{b}- \mathbf{u}_{c} - \mathbf{u}_{d} \\&=& \mathbf{B}_3
\end{eqnarray*}
with $\mathbf{B}_1$ and $\mathbf{B}_2$ exchanged. The same analysis carries through for all tetrahedra, and exchanging $S_A$ and $S_B$ everywhere on the lattice is equivalent to exchanging the coarse grained fields $\tilde{\mathbf{B}}_1$ and $\tilde{\mathbf{B}}_2$. This gives the first of the symmetries listed in Eq.~(\ref{eq:sym1}); the others can be derived in an analogous manner. 

Let us pause to consider the implications of Eq.~(\ref{eq:sym1}). The first three equations require a symmetry under exchanging any of the $\mathbf{B}_{\alpha}$ fields; this justifies having the same coefficient in front of all three quadratic, diagonal terms in the free energy Eq.~(\ref{eq:MaxwellFreeEnergy}). The last three equations forbid quadratic terms that are off-diagonal in the fields. For example, a term like $\tilde{\mathbf{B}}_1\tilde{\mathbf{B}_2}$ is not a symmetry under $\{\tilde{\mathbf{B}}_1, \tilde{\mathbf{B}}_2, \tilde{\mathbf{B}}_3\} \rightarrow \{-\tilde{\mathbf{B}}_1, \tilde{\mathbf{B}}_2, -\tilde{\mathbf{B}}_3\}$. At this point quadratic terms like $\tilde{B}_1^x \tilde{B}_1^y$ can still be added, though we will soon show that these are forbidden by spatial symmetries.

Having considered transformations of the ``internal'' spin degrees of freedom, we now turn to lattice transformations. The space group $Fd\bar{3}m$ of the pyrochlore lattice consists of the 24 element tetrahedral point group $\bar{4}3m$ and a further 24 non-symmorphic elements, involving a combination of rotations or reflections with translation by half a lattice vector. For the pyrochlore dressed with Potts spins, the space group elements transform one ground state configuration into another. 

The $\mathbf{B}_{\alpha}$, are lattice vector fields whose transformation under the space group elements (like rotations $R$) derives from the transformation of the lattice bond vectors $\mathbf{u}$. For example, the field $\mathbf{B}_1$ transforms as:
\begin{eqnarray*}
\mathbf{B}_1^i(\mathbf{r}) &=& \sum\limits_{\kappa =1}^{4}S_\kappa^x(\mathbf{r}) \mathbf{u}_\kappa^i \\
&\rightarrow& \sum\limits_{\kappa =1}^{4}S_\kappa^x(R \mathbf{r}) R^{ij} \mathbf{u}_\kappa^j \\
&\equiv& R^{ij} \mathbf{B}_1^j (R\mathbf{r}).
\end{eqnarray*}
Since the free energy involves an integral over all space, the change in the spatial location of the fields $(\mathbf{r} \rightarrow R\,\mathbf{r})$ can be undone by a simple change of integration variables; what matters is the transformation of the vector indices of the fields. Microscopically, the transformation of the vector indices derives entirely from a permutation of sublattice indices i.e. a spin belonging to sublattice 1 of a tetrahedron at location $\mathbf{r}$ gets rotated to, say, sublattice 3 of the tetrahedron at location $R \, \mathbf{r}$. In this way, all we're concerned about is the action of the space group elements in permuting sublattice indices. The 4! permutations of the sublattice indices lead to the additional symmetries: 
\begin{eqnarray}
f(B_{\alpha}^x, B_{\alpha}^y, B_{\alpha}^z) &=& f(B_{\alpha}^y, B_{\alpha}^x, B_{\alpha}^z) \nonumber \\
&=& f(B_{\alpha}^z, B_{\alpha}^y, B_{\alpha}^x) \nonumber \\
&=& f(B_{\alpha}^x, B_{\alpha}^z, B_{\alpha}^y) \nonumber \\
&=& f(-B_{\alpha}^x, -B_{\alpha}^y, B_{\alpha}^z) \nonumber \\
&=& f(-B_{\alpha}^x, B_{\alpha}^y, -B_{\alpha}^z) \nonumber \\
&=& f(B_{\alpha}^x, -B_{\alpha}^y, -B_{\alpha}^z) 
\label{eq:sym2}
\end{eqnarray}
where $\alpha = 1,2,3$ labels the type of magnetic field and we have used a compressed notation to label the nine arguments of the free energy function. 

We should emphasize that permuting sublattices is very different from permuting spins. In the latter case, we exchange spins of types $S_A$ and $S_B$ regardless of the sublattices on which they lie; in the former, we exchange the spins living on sublattice 1 and 2 regardless of their type. As shown by Eqs.~(\ref{eq:sym1}), (\ref{eq:sym2}), exchanging spins leads to symmetries under exchanging different \textit{types} of $\mathbf{B}$ fields, while exchanging sublattices leads to symmetries under exchanging different spatial \textit{components} of a given type of field. 

As before, let us consider an example to understand the symmetries listed in Eq.~(\ref{eq:sym2}). Fix the center of one tetrahedron as the origin and rotate the lattice by $\pi$ about the axis $(0,0,z)$. (This is the $C_z$ element of the tetrahedral point group). This axis bisects the edges $\mathbf{u}_{12}  \equiv (\mathbf{u}_1 - \mathbf{u}_2)$ and $\mathbf{u}_{34}$ of the tetrahedron at the origin. The rotation by $\pi$ does a dual exchange of sublattice indices $1 \rightleftarrows 2$ and $3 \rightleftarrows 4$ on each tetrahedron (in addition to rotating the tetrahedron's center). A given tetrahedron has $\mathbf{B}_1$ fields defined by
\begin{eqnarray*}
B_1^x &=& S_1^x\, u_1^x + S_2^x\, u_2^x+ S_3^x\, u_3^x+ S_4^x\, u_4^x \\
&=& 0.25(-S_1^x + S_2^x - S_3^x + S_4^x)
\end{eqnarray*}
\begin{eqnarray*}
B_1^y &=& S_1^x\, u_1^y + S_2^x\, u_2^y+ S_3^x\, u_3^y+ S_4^x\, u_4^y\\
&=& 0.25(S_1^x - S_2^x - S_3^x + S_4^x)
\end{eqnarray*}
\begin{eqnarray*}
B_1^z &=& S_1^x\, u_1^z + S_2^x\, u_2^z+ S_3^x\, u_3^z+ S_4^x\, u_4^z\\
&=& 0.25(S_1^x + S_2^x - S_3^x - S_4^x)
\end{eqnarray*}
where $\{S_1, S_2, S_3, S_4\}$ are the spins living on sublattices $\{\mathbf{u}_1, \mathbf{u}_2,\mathbf{u}_3,\mathbf{u}_4\}$ respectively as in Appendix A. Now, exchange $\mathbf{u}_1 \rightleftarrows \mathbf{u}_2$ and $\mathbf{u}_3 \rightleftarrows \mathbf{u}_4$. The transformed field equations are: 
\begin{eqnarray*}
(B_1^x)' &=& S_1^x\, u_2^x + S_2^x\, u_1^x+ S_3^x\, u_4^x+ S_4^x\, u_3^x \\
&=& 0.25(+S_1^x - S_2^x + S_3^x - S_4^x)\\
&=& -B_1^x
\end{eqnarray*}
\begin{eqnarray*}
(B_1^y)' &=& S_1^x\, u_2^y + S_2^x\, u_1^y+ S_3^x\, u_4^y+ S_4^x\, u_3^y\\
&=& 0.25(-S_1^x + S_2^x + S_3^x - S_4^x)\\
&=& -B_1^y
\end{eqnarray*}
\begin{eqnarray*}
(B_1^z)' &=& S_1^x\, u_2^z + S_2^x\, u_1^z+ S_3^x\, u_4^z+ S_4^x\, u_3^z\\
&=& 0.25(S_1^x + S_2^x - S_3^x - S_4^x)\\
&=& B_1^z.
\end{eqnarray*}
The transformation has flipped the sign of the first two components of $\mathbf{B}_1$ corresponding to the fourth symmetry in Eq.~(\ref{eq:sym2}). Of course, exactly the same transformations carry through for $\mathbf{B}_2$ and $\mathbf{B}_3$. Also note that in 3D, the matrix for rotating by $\pi$ about the $z$ axis looks like 
\begin{equation*}
R_{\pi}^z = \left(\begin{array}{ccc} -1&0&0\\0&-1&0\\0&0&1\end{array}\right)
\end{equation*}
\\whose action is also to flip the first two vector indices of the fields it acts on. 

Finally, armed with the symmetries in Eqs.~(\ref{eq:sym1}), (\ref{eq:sym2}) it is easy to see that the simplest terms we can add to the quadratic free energy defined in Eq.~(\ref{eq:MaxwellFreeEnergy}) are \textit{cubic} in the fields, and symmetric with respect to exchanging the types of fields and the $x, y, z$ components of each field:
\begin{equation}
F = F_{quad} + \int d^3r \, (B^1_xB^2_yB^3_z + B^1_yB^2_xB^3_z + \mbox{permutations}).
\label{eq:modFreeEnergy}
\end{equation} 
The added terms are cubic in the gradient of the vector potential and thus irrelevant under a renormalization-group analysis for determining the long-wavelength correlations of the fields. This confirms the stability of the quadratic, diagonal free energy used in the bulk of our analysis. In future work, it would be interesting to explicitly derive the perturbative corrections to the correlations stemming from Eq.~(\ref{eq:modFreeEnergy}).  

\end{appendix}

\bibliography{Coulomb2}

\end{document}